\documentclass[aps,showpacs,prl,reprint,superscriptaddress,groupedaddress,longbibliography]{revtex4-1}

\usepackage[colorlinks=true,linkcolor=blue,citecolor=blue,urlcolor=blue]{hyperref}

\usepackage{hyperref}
\usepackage{setspace} 
\usepackage{graphicx}
\usepackage{amsmath}
\usepackage{color}
\usepackage{amsmath}
\usepackage{amssymb}
\usepackage{verbatim}
\usepackage{latexsym}
\usepackage{enumerate} 
\usepackage{bm} 

\begin{document}

\title{Giant electron-phonon interactions in molecular crystals and the
  importance of non-quadratic coupling}

\author{Bartomeu Monserrat}
\email{bm418@cam.ac.uk}
\affiliation{TCM Group, Cavendish Laboratory, University of Cambridge,
  J.\ J.\ Thomson Avenue, Cambridge CB3 0HE, United Kingdom}

\author{Edgar A. Engel}
\affiliation{TCM Group, Cavendish Laboratory, University of Cambridge,
  J.\ J.\ Thomson Avenue, Cambridge CB3 0HE, United Kingdom}

\author{Richard J.\ Needs}
\affiliation{TCM Group, Cavendish Laboratory, University of Cambridge,
  J.\ J.\ Thomson Avenue, Cambridge CB3 0HE, United Kingdom}

\date{\today}

\begin{abstract}
  We investigate electron-phonon coupling in the molecular crystals
  CH$_4$, NH$_3$, H$_2$O, and HF, using first-principles quantum
  mechanical calculations. We find vibrational corrections to the
  electronic band gaps at zero temperature of $-1.97$~eV, $-1.01$~eV,
  $-1.52$~eV, and $-1.62$~eV, respectively, which are comparable in
  magnitude to those from electron-electron correlation effects.
  Microscopically, the strong electron-phonon coupling arises in
  roughly equal measure from the almost dispersionless high-frequency
  molecular modes and from the lower frequency lattice modes.  We also
  highlight the limitations of the widely used Allen-Heine-Cardona
  theory, 
  which gives significant discrepancies 
  compared to our more accurate treatment.
\end{abstract}


\maketitle


Electronic band gaps are crucial for studying the optical and
electrical properties of semiconductors, and are therefore central to
modern technologies such as photovoltaics and thermoelectrics. Density
functional theory
(DFT)~\cite{PhysRev.136.B864,PhysRev.140.A1133,dft_rev_mod_phys} using
local and semi-local density
functionals~\cite{PhysRevLett.45.566,PhysRevB.23.5048,PhysRevLett.77.3865}
notoriously suffers from the \textit{band gap problem} caused by the
inability of standard local and semi-local exchange-correlation
functionals to reproduce the derivative discontinuity with respect to
particle number exhibited by the exact
functional~\cite{PhysRevLett.56.2415,PhysRevB.37.10159}. This
typically leads to underestimation of band gaps. Computationally
intensive techniques such as diffusion Monte Carlo
(DMC)~\cite{RevModPhys.73.33,original_dmc_excited,silicon_dmc_excited},
\textit{GW} many-body perturbation
theory~\cite{original_gw_hedin,0034-4885-61-3-002}, and hybrid
exchange-correlation
functionals~\cite{b3lyp,pbe0,hybrid_band_gaps,hse06_functional,hse06_functional_erratum}
have been shown to greatly improve upon the accuracy of DFT band gaps,
with corrections of several eV being common in some materials.
However, even these more accurate methods assume a static lattice and
neglect the effects of atomic vibrations.

One of the most striking consequence of electron-phonon coupling in
semiconductors is the temperature dependence of electronic band
gaps~\cite{RevModPhys.77.1173}. Early
experimental~\cite{ZPhys.fesefeldt,Clark11021964,silicon_gap_tdependence,diamond_tdependence_experiment}
and
theoretical~\cite{0022-3719-9-12-013,PhysRevB.23.1495,PhysRevB.45.3376}
efforts developed understanding of the temperature dependence of band
gaps in standard semiconductors such as diamond, silicon and
germanium. Recent interest has focused on first-principles
calculations~\cite{0295-5075-10-6-011,nanotube_t_dependence,PhysRevLett.105.265501,PhysRevLett.107.255501,cannuccia_ejpb,PhysRevB.87.144302,giustino_nat_comm,elph_Si_nano,gonze_marini_elph,gonze_gw_elph,monserrat_elph_diamond_silicon},
which have mainly been used to study tetrahedral semiconductors, and
particularly diamond due to the large electron-phonon coupling
strength it exhibits. There is also great interest in electron-phonon
coupling in
nanocrystals~\cite{nanotube_t_dependence,giustino_nat_comm,elph_Si_nano},
topological
insulators~\cite{elph_topological_prl,elph_topological_prb}, and
polymers~\cite{elph_polymer}.

Systematic studies of electron-phonon coupling over a wide range of
classes of materials are currently lacking. Such studies could
elucidate the interplay between the various microscopic properties
that play central roles in electron-phonon coupling in semiconductors,
and ultimately allow us to build an understanding that could be used
to design new materials with tailored electronic and optical
properties.

We gauge the strength of electron-phonon coupling by calculating the
zero-point (ZP) vibrational correction to the electronic band gap.  As
far as we are aware, this is the first report of ZP corrections to the
band gaps of the molecular crystals studied here.  We find that the
commonly used Allen-Heine-Cardona (AHC) theory is insufficient for
studying electron-phonon coupling in the molecular crystals
considered, and therefore we include terms beyond the AHC theory using
a Monte Carlo sampling approach~\cite{giustino_nat_comm,helium}. We
find ZP corrections ranging from $-1.0$~eV in NH$_3$, to $-2.0$~eV in
CH$_4$.  These corrections are the largest observed to date, apart
from those in hydrogen and helium at extreme
pressures~\cite{ceperley_h_elph_coupling,helium}. The large ZP
corrections arising from electron-phonon coupling demonstrate that
studies of the electronic structure of semiconductors should include
the effects of atomic vibrations even at zero temperature.




The ZP corrected band gap $E_{\mathrm{g}}$ may be written as the expectation value 
\begin{equation}
  E_{\mathrm{g}}\!=\!\langle\Phi(\mathbf{q})|E_{\mathrm{g}}(\mathbf{q})|\Phi(\mathbf{q})\rangle, \label{eq:exp_value}
\end{equation}
where $E_{\mathrm{g}}(\mathbf{q})$ is the value of the gap with atomic
positions $\mathbf{q}$ (a $3N$-dimensional vector in the vibrational
phase space, which we describe in a basis of harmonic normal
mode coordinates), and $|\Phi\rangle$ is the Gaussian vibrational ground state wave
function. 
Equation~(\ref{eq:exp_value}) is based on adiabatic decoupling of the
electronic and nuclear degrees of freedom, which has recently been
explored by Patrick and Giustino \cite{patrick_molecule_solid_2014},
and shown to provide a semiclassical approximation to the ZP
correction of band gaps.

In this work we evaluate the band gap correction of
Eq.~(\ref{eq:exp_value}) using two different methods: the quadratic
approximation and Monte Carlo sampling. In the quadratic
approximation, the electronic band gap energy $E_{\mathrm{g}}$ is
expanded about the equilibrium position in terms of harmonic
vibrational mode amplitudes $q_{n\mathbf{k}}$,
where $\mathbf{k}$ is the vibrational Brillouin zone wavevector and 
$n$ is the branch index.  We work exclusively within the harmonic
vibrational approximation, so that the wave function is symmetric and
odd terms vanish in the expectation value:
\begin{equation}
  E_{\mathrm{g}}\!=\!E_{\mathrm{g}}(\mathbf{0})+\sum_{n,\mathbf{k}}a_{n\mathbf{k}}^{(2)}\langle\Phi_{n\mathbf{k}}|q_{n\mathbf{k}}^2|\Phi_{n\mathbf{k}}\rangle + \mathcal{O}(q^4), \label{eq:exp_value_quad}
\end{equation}
where $a_{n\mathbf{k}}^{(2)}$ are the diagonal quadratic expansion
coefficients.  Eq.~(\ref{eq:exp_value_quad}) is equivalent to the AHC
theory including off-diagonal Debye-Waller
terms~\cite{gonze_off_diagonal}.

\begin{figure}
\centering
\includegraphics[scale=0.33]{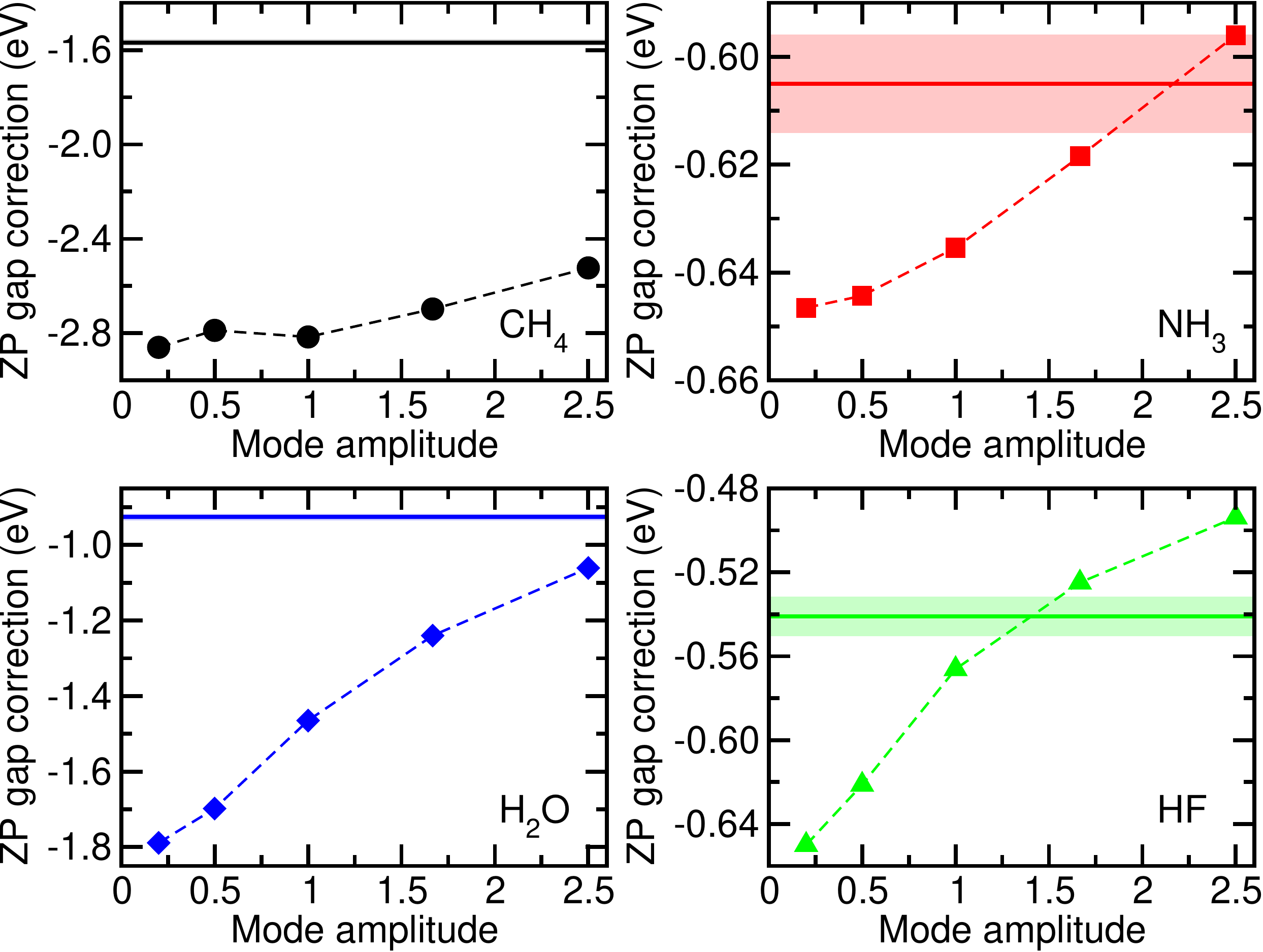}
\caption{(color online) ZP correction to the electronic band gaps for
  the molecular crystals CH$_4$, NH$_3$, H$_2$O, and HF. The solid
  horizontal lines correspond to accurate Monte Carlo sampling, and
  the light bands represent their statistical uncertainty. The symbols
  correspond to results within the quadratic approximation, and the
  ``Mode amplitude'' refers to the amplitude at which the frozen
  phonon calculations have been performed. The mode amplitudes have
  units of $\sqrt{\langle q^2\rangle}=1/\sqrt{2\omega}$.}
\label{fig:convergence}
\end{figure}

We gauge the strength of the electron-phonon coupling by calculating
the ZP correction to the band gap. 
We calculate the couplings $a^{(2)}_{n\mathbf{k}}$ using a
frozen phonon method, displacing the atoms by $\Delta q_{n\mathbf{k}}$
along harmonic vibrational modes in the positive and negative
directions, and setting
$  a_{n\mathbf{k}}^{(2)}=(E_{\mathrm{g}}(\Delta q_{n\mathbf{k}})+E_{\mathrm{g}}(-\Delta q_{n\mathbf{k}}))/2\Delta q^2_{n\mathbf{k}}$. 
We have found that $a_{n\mathbf{k}}^{(2)}$ is sensitive to the
vibrational amplitude used ($\Delta q_{n\mathbf{k}}$), as shown in
Fig.~\ref{fig:convergence} and discussed below. This sensitivity is a
signature of a strongly non-quadratic dependence of $E_{\mathrm{g}}$ on
the normal mode amplitude $q_{n\mathbf{k}}$.

We also evaluate Eq.~(\ref{eq:exp_value}) using Monte Carlo sampling,
with atomic configurations randomly drawn from the vibrational
density. A Monte Carlo approach has an associated statistical
uncertainty that can be reduced by using a large number of sampling
points. The Monte Carlo approach allows us to include all higher-order
terms neglected in Eq.~(\ref{eq:exp_value_quad}). Here we use both
approaches, and assess the validity of the quadratic approximation by
comparing the results of the two methods. This comparison highlights
the importance of higher-order terms and the inadequacy of the
quadratic expansion for the molecular crystals considered. The only
other known example of the failure of AHC theory has arisen in helium
at terapascal pressures~\cite{helium}, and our work emphasizes that
AHC theory can also be inadequate at ambient pressure. It is still
valuable to use both approaches, as the quadratic expansion provides
direct access to the microscopic physics, and therefore allows us to
build a physical understanding of the origin of the strong
electron-phonon coupling in these materials. Finally, we note that
Brooks' theorem~\cite{ZPBAllen} breaks down if non-quadratic terms are
important.

All our calculations have been performed using DFT and the
pseudopotential plane-wave method as implemented in the {\sc castep}
package~\cite{castep}. We have used ultrasoft ``on the fly''
pseudopotentials to describe the electron-ion
interaction~\cite{PhysRevB.41.7892}. We use an energy cutoff of
$800$~eV and a Monkhorst-Pack~\cite{PhysRevB.13.5188} Brillouin zone
sampling of spacing $2\pi\times0.03$~\AA$^{-1}$, that together lead to
convergence of the difference between frozen-phonon structures to
better than $10^{-4}$~eV for the total energy per atom,
$10^{-3}$~eV/\AA\@ for the forces on all atoms, $10^{-2}$~GPa for the
components of the stress tensor, and $10^{-4}$~eV for the band
gap. Unless otherwise stated, we report results using the PBE
functional corrected with the TS scheme which describes dispersion
interactions~\cite{ts_vdW}, as equilibrium volumes using this
functional agree best with experimental volumes. Results obtained
with the PBE~\cite{PhysRevLett.77.3865} functional and the
G06~\cite{go6_vdW} dispersion corrected functional are detailed in the
Supplementary Material~\cite{supplementary_molec}. The ZP correction arising from electron-phonon
coupling has some dependence on the functional and geometry used, but
the central idea presented in this paper -- the large strength of
electron-phonon coupling -- is independent of the choice of
functional.

Recent \textit{GW} calculations have shown that electron-electron
correlation plays an important role in the effects of electron-phonon
coupling on the band gap of diamond and GaAs~\cite{gonze_gw_elph},
increasing the ZP correction within semi-local DFT by about $40\%$.  We
have estimated the importance of electron-electron correlation by
performing calculations with the HSE06
functional~\cite{hse06_functional,hse06_functional_erratum}, which has
been shown to provide improved static lattice band gaps within DFT. We
find somewhat smaller corrections than those reported for diamond and
GaAs using \textit{GW} but, as in Ref.~\cite{gonze_gw_elph}, these
corrections increase the ZP renormalisation (details are provided in
the Supplementary Material~\cite{supplementary_molec}).


The quadratic approximation of Eq.~(\ref{eq:exp_value_quad}) is a
computationally efficient approach for evaluating the ZP correction to
the band gap. Furthermore, it provides access to the microscopic
origins of the electron-phonon coupling strength, as each vibrational
normal mode is treated separately.
The importance of higher order terms in the expansion of
Eq.~(\ref{eq:exp_value_quad}) is less well-understood.  These terms
have been found to be crucial for the electron-phonon coupling
strength in helium under terapascal conditions~\cite{helium}, but to
be unimportant for silicon at ambient
pressure~\cite{patrick_molecule_solid_2014}.  The question arises of
how important they may be for the molecular crystals of interest here.

\begin{figure}
\centering
\includegraphics[scale=0.4]{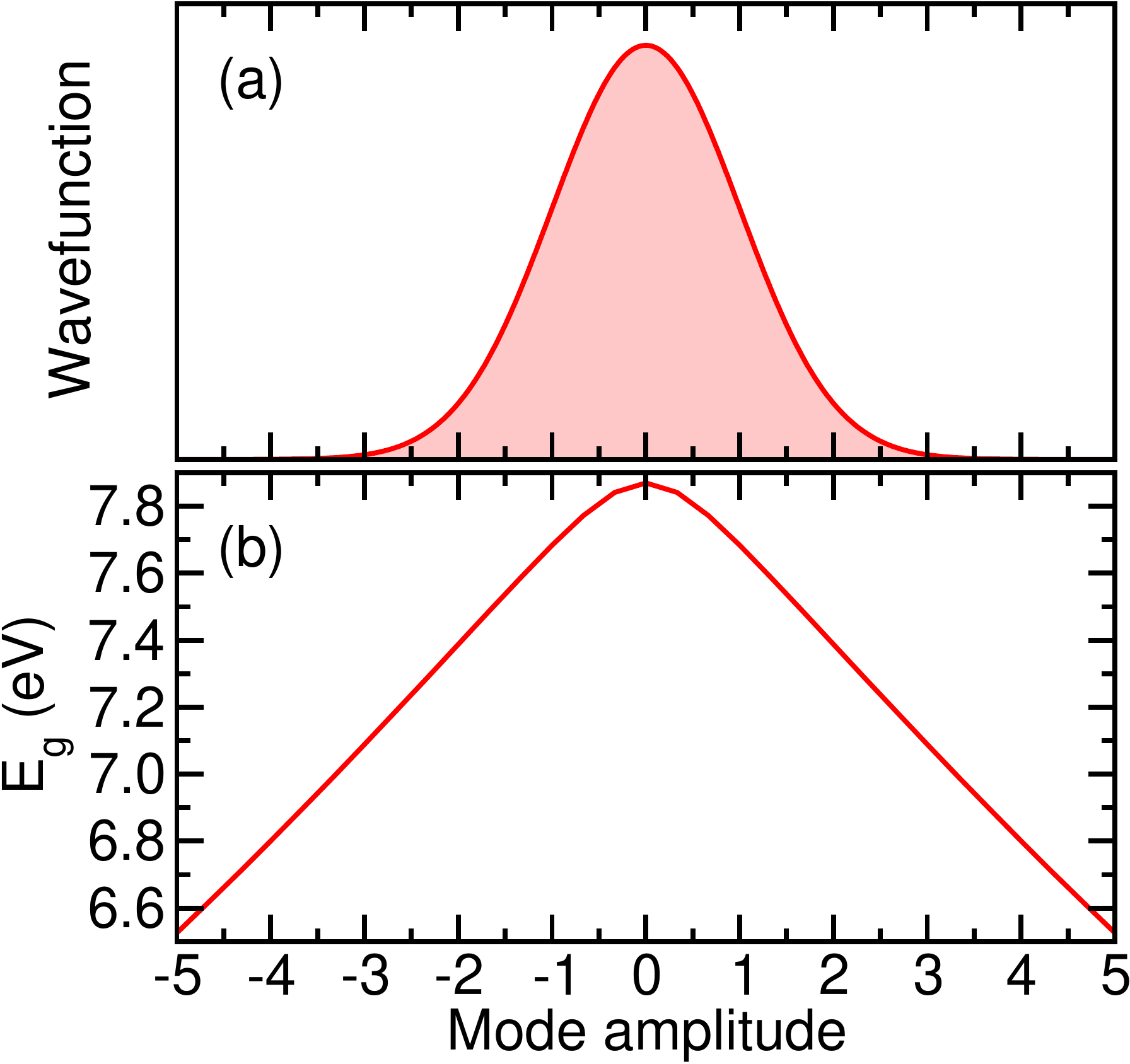}
\caption{(a) Wave function and (b) band gap $E_{\mathrm{g}}$ as a
  function of normal mode amplitude (in units of $1/\sqrt{2\omega}$)
  for a high-energy vibrational molecular mode of the HF crystal.}
\label{fig:nonquadratic}
\end{figure}

In Fig.~\ref{fig:convergence} we compare the ZP correction to the band
gaps of CH$_4$, NH$_3$, H$_2$O, and HF, evaluated using the quadratic
expansion and Monte Carlo sampling. The horizontal axes show the
vibrational mode amplitude at which the band gap is calculated in
order to evaluate the couplings
$a^{(2)}_{n\mathbf{k}}$. 
The results reported in Fig.\ \ref{fig:convergence} correspond to
primitive cells ($\Gamma$-point sampling of the vibrational Brillouin
zone), which are sufficient to illustrate the discrepancies between
the two methods. Note that all other results in the paper are reported
for larger simulation cells.  We observe a strong dependence on the
normal mode amplitude of the ZP correction evaluated within the
quadratic approximation. It is clear that higher order terms in
Eq.~(\ref{eq:exp_value_quad}) are large, and should be included to
obtain accurate results. We show an example of non-quadratic
dependence in Fig.~\ref{fig:nonquadratic}, corresponding to a normal
mode describing a high-energy molecular vibration in HF. The
dependence of $E_{\mathrm{g}}$ on mode amplitude has a near-linear
behaviour at normal mode amplitudes larger than about
$1/\sqrt{2\omega}$, and this non-quadratic region is clearly important
given the width of the vibrational wave function shown.

Higher order terms in Eq.~(\ref{eq:exp_value_quad}) have largely been
neglected in the literature, and the results presented here and in
Ref.~\onlinecite{helium} suggest that future studies should carefully
assess their importance. Here we report accurate numerical results
from Monte Carlo evaluations and results from the quadratic expansion
to investigate the microscopic origins of the electron-phonon
coupling. Further comparisons of the quadratic and Monte Carlo
approaches are given in the Supplementary Material~\cite{supplementary_molec}.

\begin{figure*}
\centering
\includegraphics[scale=0.85]{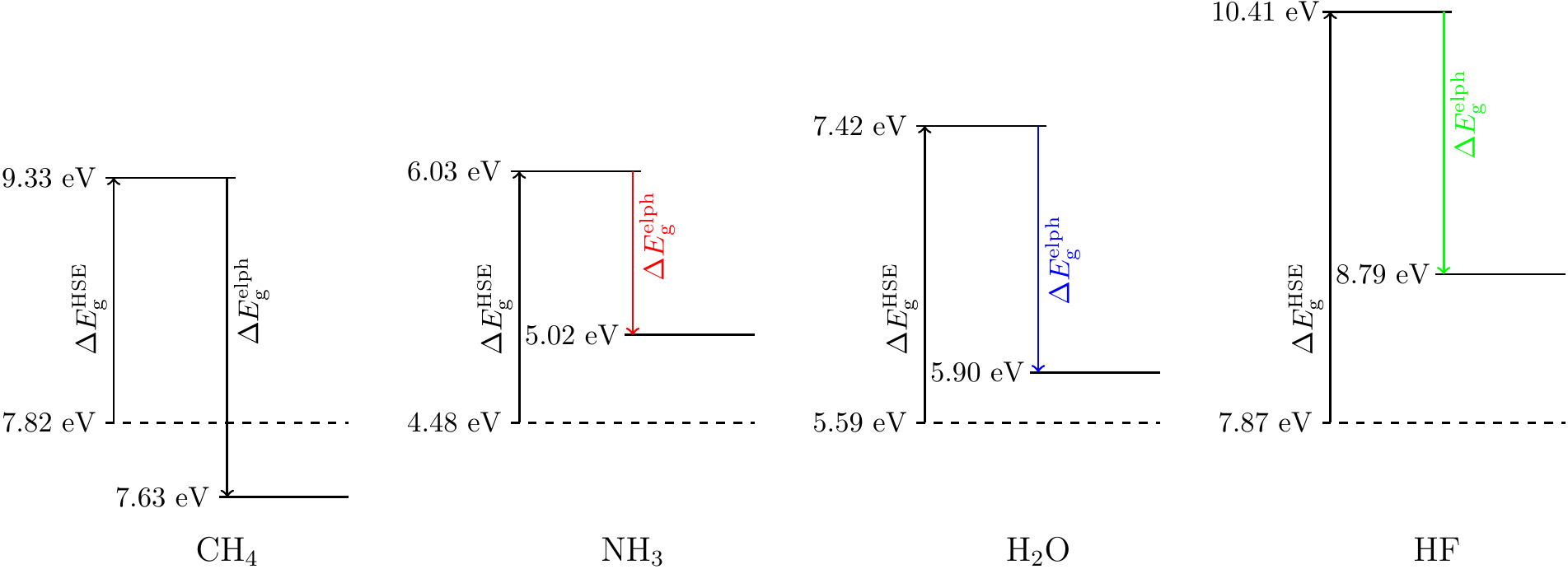}
\caption{(color online) Band gap energies at the static lattice level
  with semi-local functionals (dashed baselines), the HSE06 correction
  $\Delta E_{\mathrm{g}}^{\mathrm{HSE}}$, and the ZP electron-phonon
  coupling correction $\Delta E_{\mathrm{g}}^{\mathrm{elph}}$.}
\label{fig:zp}
\end{figure*}

In Fig.~\ref{fig:zp} we show schematic diagrams of the band gaps of
the molecular crystals CH$_4$, NH$_3$, H$_2$O and HF. The dashed
baseline corresponds to the static lattice band gap evaluated using the
PBE-TS functional. The corrections arising from the screened-exchange
HSE06 functional are indicated by the arrows labelled $\Delta
E_{\mathrm{g}}^{\mathrm{HSE}}$, and the corrections induced by
electron-phonon coupling at zero temperature are indicated by the
coloured arrows labelled by $\Delta
E_{\mathrm{g}}^{\mathrm{elph}}$. In each case the ZP correction to the
band gap has been calculated with a large supercell using the PBE or
PBE-TS functional. The Monte Carlo sampling approach has been used,
achieving a statistical uncertainty smaller than $0.02$~eV in all
cases.  The ZP corrections obtained using the HSE06 functional are
similar to those obtained with PBE-TS (see discussion in Supplementary
Material~\cite{supplementary_molec}).  It is worth mentioning that the static HSE06 gap for ice
is significantly smaller than many-body perturbation theory estimates,
which are in the range
$9.2$--$10.1$~eV~\cite{ice_gw_gap1,ice_gw_gap2}.

We observe very large vibrational ZP corrections to the band gaps 
across the four systems studied of above $1$~eV. The corrections are 
comparable to 
those arising from the use of the HSE06 screened exchange functional
when compared with the semi-local functionals. These giant ZP corrections
are the largest found to date in any system at ambient conditions, and
are about twice the ZP correction in diamond. We note that hydrogen
and helium solids under extreme pressures have been reported to
exhibit even larger ZP
corrections~\cite{ceperley_h_elph_coupling,helium}.
First-principles molecular dynamics calculations have also shown a
large influence of thermal motion on the band gap of liquid
water~\cite{galli_water_gap_aimd}, of a size comparable to the
electron-phonon effects reported in this work.

To investigate the microscopic origins of the strong electron-phonon
coupling, we evaluate the contribution from individual normal
vibration modes to the ZP correction using the quadratic method.  

In CH$_4$, the high frequency and virtually dispersionless molecular
modes contribute about $69$\% of the overall ZP correction to the band
gaps, and of
these, the low energy twisting modes dominate, contributing as much as
the combination of the higher energy stretching and wagging modes. The
crystal modes contribute the remaining $31$\% to the overall ZP
correction.  In NH$_3$, the molecular modes contribute about $55$\% of
the overall ZP correction, with the low energy wagging modes
dominating, followed by the high energy stretching modes, and finally
the intermediate energy scissoring modes. The crystal modes contribute
$45$\% to the overall ZP correction, and the dominant modes involve
vibrations of hydrogen atoms which contribute $37$\% of the overall ZP
correction.  For H$_2$O, the contributions from molecular and crystal
modes are $46$\% and $54$\%, respectively, and for HF they are $51$\%
and $49$\%.

\begin{figure}
\centering
\includegraphics[scale=1.1]{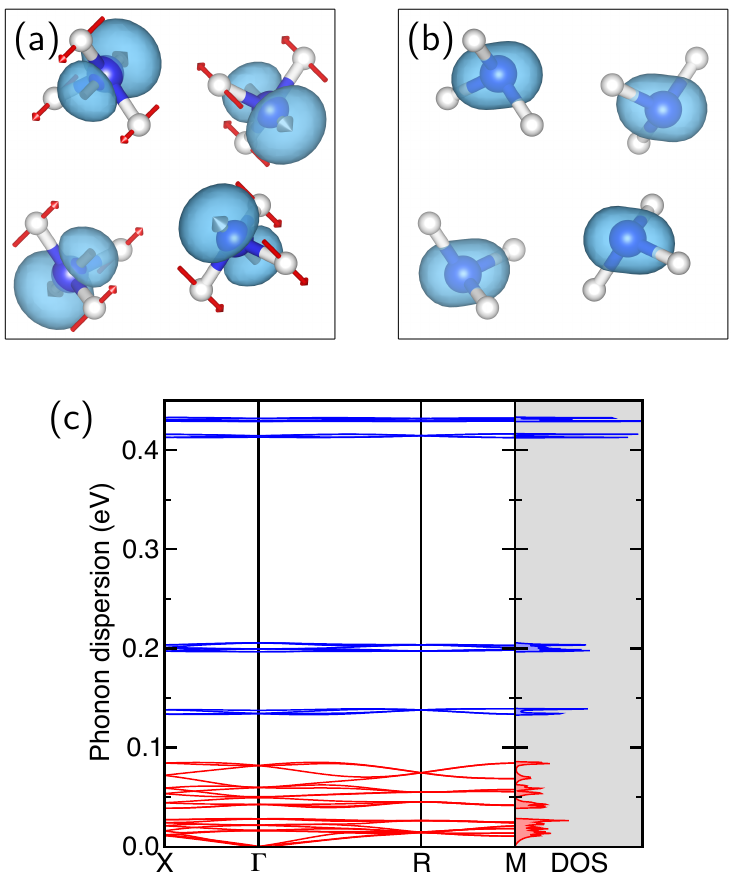}
\caption{(color online) Electronic charge density in crystalline
  NH$_3$ corresponding to (a) the VBM and (b) the CBM. (c)~Phonon
  dispersion (left) and DOS (right) of NH$_3$, where the low-energy
  lattice modes are shown in red, and the high-energy molecular modes
  are shown in blue.}
\label{fig:orbitals}
\end{figure}

The microscopic origin of the strength of electron-phonon coupling can
be understood by relating atomic distortions to changes in the charge
density associated with the valence band maximum (VBM) and the
conduction band minimum (CBM).  As an example, we consider the NH$_3$
molecular crystal.  In Fig.~\ref{fig:orbitals} we show the charge
density of the VBM and CBM for the NH$_3$ molecular crystal. Both
electronic states couple strongly to atomic vibrations. The phonon
dispersion and density of states (DOS) of NH$_3$ are also shown in
Fig.~\ref{fig:orbitals}, where the low-energy lattice modes are shown
in red, and the molecular quasi-dispersionless high-energy modes are
shown in blue. The VBM charge density distorts significantly under
molecular vibrations, in particular the wagging modes in which the
H-atoms move 
in the direction of the charge density of the VBM (indicated as arrows
in Fig.~\ref{fig:orbitals}), explaining the strong coupling to it.
These wagging modes exhibit some of the strongest electron-phonon
coupling in NH$_3$. The charge density of the CBM is mostly affected
by the low-energy lattice vibrational modes that involve collective
H-motion.

For comparison, we have also calculated the ZP contribution to the
band gap of the isolated molecules. As observed for their crystalline
counterparts, the molecular calculations also exhibit a strong
non-quadratic behaviour, and we calculate Monte Carlo ZP corrections
of $-0.39$~eV for CH$_4$, $-1.32$~eV for NH$_3$, $-0.48$~eV for
H$_2$O, and $-0.04$~eV for HF. Apart from NH$_3$, the molecular
correction is significantly smaller in the molecules than in the
crystals.



In conclusion, recent efforts to understand the effects of
electron-phonon coupling in semiconductors have highlighted their
importance for calculating accurate band structures using
first-principles methods. We have reported the largest ZP corrections
to band gaps found to date in solids under standard conditions of
pressure, and we hope to motivate further studies of a wider range of
systems.  We have shown that the commonly used quadratic approximation
fails for the description of the ZP correction to the band gap of
hydrogen-rich molecular crystals. It would be interesting to assess
the importance of dynamical effects on the calculated ZP corrections,
but this is beyond the scope of the present work.

\textit{GW} calculations are often used to correct band gaps obtained
with semi-local density functionals.  \textit{GW} results are
typically benchmarked against experimental data, but our results
indicate that such comparisons could incur errors as large as $1$~eV,
and therefore care should be taken when assessing the accuracy of such
calculations.  The large ZP corrections found in molecular crystals
could also be present in pharmaceutical drugs, polymers, or
biological molecules. 

Finally, we have highlighted the limits of the quadratic or AHC
approximation for calculating band gap corrections arising from
electron-phonon coupling. Most studies of these effects are based on
the quadratic approximation, and our results call for a careful
assessment of its validity over a wide range of materials.

B.M.\ acknowledges Robinson College, Cambridge, and the Cambridge
Philosophical Society for a Henslow Research Fellowship. E.A.E.\ and
R.J.N.\ acknowledge financial support from the Engineering and
Physical Sciences Research Council (EPSRC) of the UK
[EP/K013688/1]. The calculations were performed on the Cambridge High
Performance Computing Service facility and the Archer facility of the
UK's national high-performance computing service (for which access was
obtained via the UKCP consortium [EP/K013564/1]).

\bibliography{/Users/bartomeumonserrat/Documents/research/papers/references/anharmonic}

\end{document}